\documentstyle[epsfig,12pt]{article}  
\newcommand{\beq}{\begin{equation}}
\newcommand{\eeq}{\end{equation}}
\newcommand{\bea}{\vspace{0.25cm}\begin{eqnarray}}
\newcommand{\eea}{\end{eqnarray}}

\newcommand{\r}{\mbox{{\boldmath
$\rho$}}}

\newcommand{\pb}{{{\bf p}}}

\newcommand{\bb}{{{\bf b}}}

\def\lsim{\mathrel{\rlap{\lower4pt\hbox{\hskip1pt$\sim$}}
    \raise1pt\hbox{$<$}}}         
\def\gsim{\mathrel{\rlap{\lower4pt\hbox{\hskip1pt$\sim$}}
    \raise1pt\hbox{$>$}}}         
\begin{document}
\vspace*{-2cm}
 
\bigskip

\begin{center}

\renewcommand{\thefootnote}{\fnsymbol{footnote}}

  {\Large\bf
Jet quenching with running coupling 
including radiative and collisional energy losses
\\
\vspace{.7cm}
  }
\renewcommand{\thefootnote}{\arabic{footnote}}
\medskip
{\large
  B.G.~Zakharov
  \bigskip
  \\
  }
{\it
 L.D.~Landau Institute for Theoretical Physics,
        GSP-1, 117940,\\ Kosygina Str. 2, 117334 Moscow, Russia
\vspace{1.7cm}\\}

  {\bf
  Abstract}
\end{center}
{
\baselineskip=9pt
We calculate the nuclear modification factor for
RHIC and LHC conditions accounting for 
the radiative and collisional parton 
energy loss with the running coupling constant.
We find that the RHIC data can be explained both in the scenario with 
the chemically
equilibrium quark-gluon plasma and purely gluonic plasma 
with slightly different thermal suppression of the coupling constant. 
The role of the parton energy gain due to gluon absorption is also 
investigated.
Our results show that the energy gain gives negligible  effect. 
\vspace{.5cm}
\\
}

\noindent{\bf 1.} 
It is widely believed that suppression of the high-$p_{T}$ hadrons in 
$AA$-collisions 
(jet quenching (JQ)) observed at RHIC (for a review, 
see \cite{RHIC_data}) is dominated by 
the induced gluon emission \cite{BDMPS,LCPI,BSZ,W1,GLV1,AMY} in the hot quark-gluon
plasma (QGP) produced at the initial stage of $AA$-collisions.
There are currently considerable theoretical efforts 
in the development of the quantitative methods for computation of  
JQ \cite{BDMS_RAA,Eskola,Gale,Armesto,Zapp,Lokhtin} which
can be used for the tomographic analysis of the 
QGP.  
In the present paper we study JQ
using the light-cone path 
integral (LCPI) approach to the radiative energy loss \cite{LCPI,BSZ}.  
In this formalism the probability
of gluon emission is expressed through the Green's function
of a two-dimensional Schr\"odinger equation with an imaginary 
potential. This approach 
has not the restrictions on the applicability of 
the BDMPS approach \cite{BDMPS} (valid only for massless 
partons in the limit of strong Landau-Pomeranchuk-Migdal effect) and 
the GLV formalism \cite{GLV1} (applicable only to a thin plasma
in the regime of small Landau-Pomeranchuk-Migdal suppression).
We perform the calculations with accurate treatment of the Coulomb effects.
If one neglects these effects the gluon spectrum can be 
expressed in terms of the oscillator Green's function and the medium
may be characterized by the well-known transport coefficient $\hat{q}$ 
\cite{BDMPS,BSZ}.
However, the oscillator approximation can lead to uncontrolled errors
since it gives 
a physically absurd prediction that for massless partons 
the dominating $N=1$ rescattering contribution vanishes \cite{Z_OA,AZZ}. 
Besides the radiation energy loss we include 
the collisional energy loss. Both the contributions
are calculated with the running coupling constant.
Also, we investigate the impact of the parton
energy gain due to gluon absorption from the QGP on JQ.

We calculate the nuclear modification factor $R_{AA}$,
which characterizes JQ, 
accounting for  the fluctuations of the parton path lengths
in the QGP. In the treatment of multiple gluon emission
we use a new method which takes into account time ordering of the DGLAP and
the induced radiation stages.
We compare the theoretical results with the data obtained at RHIC by 
the PHENIX Collaboration\cite{PHENIX08}
and give prediction for LHC. Our principle purpose in comparing with the 
RHIC data is to understand whether the observed JQ is consistent with
the entropy of the QGP required by the hydrodynamical simulations of the
$AA$-collisions for reproducing the observed particle multiplicities.
Our results show that JQ and particle multiplicities
can be naturally reconciled.
Contrary to the conclusion of Ref. \cite{BMueller}
that the observed at RHIC JQ is consistent only with purely gluonic plasma,
we find that the scenario with the chemically equilibrium QGP
is also possible. A good description of the JQ RHIC data can 
be obtained in this scenario with the thermal
suppression of the coupling constant qualitatively consistent with the 
lattice results.

\vspace{.2cm}
\noindent{\bf 2.}
As usual we define the nuclear modification factor for $AA$-collisions
as
\beq
R_{AA}(b)=\frac{{dN(A+A\rightarrow h+X)}/{d\pb_{T}dy}}
{T_{AA}(b){d\sigma(N+N\rightarrow h+X)}/{d\pb_{T}dy}}\,,
\label{eq:10}
\eeq
where $\pb_{T}$ is the hadron transverse momentum, $y$ is rapidity (we
consider the central region $y=0$), 
$b$ is the impact parameter, 
$T_{AA}(b)=\int d\r T_{A}(\r) T_{A}(\r-\bb)$, $T_{A}$ is the nucleus 
profile function. The differential yield for high-$p_{T}$ hadron production in 
$AA$-collision can be written in the form 
\beq
\frac{dN(A+A\rightarrow h+X)}{d\pb_{T} dy}=\int d\r T_{A}(\r)T_{B}(\r-\bb)
\frac{d\sigma_{m}(N+N\rightarrow h+X)}{d\pb_{T} dy}\,,
\label{eq:20}
\eeq
where ${d\sigma_{m}(N+N\rightarrow h+X)}/{d\pb_{T} dy}$ is the medium-modified
cross section for the $N+N\rightarrow h+X$ process.
In analogy to the ordinary pQCD formula, we write it in the form
\beq
\frac{d\sigma_{m}(N+N\rightarrow h+X)}{d\pb_{T} dy}=
\sum_{i}\int_{0}^{1} \frac{dz}{z^{2}}
D_{h/i}^{m}(z, Q)
\frac{d\sigma(N+N\rightarrow i+X)}{d\pb_{T}^{i} dy}\,.
\label{eq:30}
\eeq
Here $\pb_{T}^{i}=\pb_{T}/z$ is the parton transverse momentum, 
$D_{h/i}^{m}$ is the medium-modified fragmentation function (FF)
for transition of the parton $i$ to the observed hadron $h$, and
${d\sigma(N+N\rightarrow i+X)}/{d\pb_{T}^{i} dy}$ is the ordinary
hard cross section.
For the parton virtuality scale $Q$ we take the parton transverse
momentum $p^{i}_{T}$.
We assume that hadronization
of the fast partons occurs after escaping from the QGP.
This hadronization process should be described by the FFs
at relatively small fragmentation scale, $\mu_h$.
Indeed, from the uncertainty relation $\Delta E\Delta t\gsim 1$
one can obtain for the $L$ dependence of the parton virtuality 
$Q^{2}(L)\sim \max{(Q/L,Q_0^{2})}$, where we have introduced some 
minimal nonperturbative scale
$Q_{0}\sim 1-2$ GeV.
For RHIC and LHC conditions the size of the QGP
is quite large ($\gsim R_A$, where $R_A$ is the nucleus radius),
and from the above formula one sees that for partons with energy 
$E\lsim 100$ GeV the hadronization of the final partons
may be described by the FFs at the scale $\mu_{h}\sim Q_{0}$.
Then we can write
\beq
D_{h/i}^{m}(z,Q)\approx\int_{z}^{1} \frac{dz'}{z'}D_{h/j}(z/z',Q_{0})
D_{j/i}^{m}(z',Q_{0},Q)\,,
\label{eq:40}
\eeq
where 
$D_{h/j}(z,Q_{0})$ is the FF in vacuum, and
$D_{j/i}^{m}(z',Q_{0},Q)$ is the medium-modified FF
for transition of the initial parton $i$ with virtuality $Q$
to the parton $j$ with the virtuality $Q_{0}$.
Presently there is no a systematic 
method for calculation of the medium-modified FFs
which treats on an even footing the DGLAP and induced radiation 
processes. In the present paper we use the picture based on the time
ordering of the DGLAP and the induced radiation stages which
should be a reasonable approximation for not very high parton energies,
say, $E\lsim 100$ GeV. It uses the fact that at such energies the typical
length/time scale of the DGLAP
stage is smaller than the longitudinal scale of the induced radiation
stage. The gluon emission scale for the DGLAP stage can be estimated
using the gluon formation length 
$l_{F}(x,k_{T}^{2})\sim 2Ex(1-x)/(k_{T}^{2}+\epsilon^{2})$,
where $x$ is the gluon fractional longitudinal momentum,
and $\epsilon$ in terms of the effective parton masses reads
$\epsilon^{2}=m_{q}^{2}x^{2}+m_{g}^{2}(1-x)$. 
Using the vacuum spectrum of the gluon emission from a quark
\beq
\frac{dN}{dk_{T}^{2}dx}=\frac{C_{F}\alpha_{s}(k_{T}^{2})}{\pi x}
\left(1-x+x^{2}/2)\right)\frac{k_{T}^{2}}{(k_{T}^{2}+\epsilon^{2})^{2}}\,
\label{eq:50}
\eeq
one can obtain for the typical formation length $\bar{l}_{F}\sim 0.3-1$ fm 
for $E\lsim 100$ GeV (if one takes $m_{q}\sim 0.3$ GeV and 
$m_{g}\sim 0.75$ GeV \cite{NZ_HERA}). This estimate is obtained in the one 
gluon approximation. However, it should be qualitatively correct since 
in the energy interval of interest the number emitted gluons is small
$\bar{N}_{g}\lsim 2$, and the first hardest gluon dominates 
the DGLAP energy loss.
Thus we see that the DGLAP time scale is about the formation time for 
the QGP, $\tau_{0}\sim 0.5-1$ fm.
Since the induced radiation is dominated by the 
distances
from $L\sim \tau_{0}$ up to $L\sim R_{A}$ one can neglect the interference 
between the DGLAP and the induced radiation stages. In this approximation
we can write
\beq
D_{j/i}^{m}(z,Q_{0},Q)=\int_{z}^{1} \frac{dz'}{z'}D_{j/l}^{ind}(z/z',E_{l})
D_{l/i}^{DGLAP}(z',Q_{0},Q)\,,
\label{eq:60}
\eeq
where $E_{l}=Qz'$,
$D_{j/l}^{ind}$ is the induced radiation FF
(it depends on the parton energy $E$, but not the virtuality), and 
$D_{l/i}^{DGLAP}$ is the DGLAP partonic FF.
In numerical calculations the DGLAP FFs
have been evaluated with the help of the PYTHIA event generator 
\cite{PYTHIA}.
 
The induced radiation FFs have been calculated making use the probability
distribution of the $1\rightarrow 2$ partonic processes obtained in the 
LCPI approach. We have taken into account only the processes with gluon
emission, and the process $g\rightarrow q\bar{q}$ which gives a small
contribution has been neglected. For calculation the one gluon emission
distribution we use the method  elaborated in \cite{Z04_RAA}.
To calculate the $D_{j/l}^{ind}$
one needs to take into account the multiple gluon emission. Unfortunately, 
up to now, there is no an accurate method of incorporating
the multiple gluon emission. 
We follow the analysis \cite{BDMS_RAA} and 
employ the Landau method
developed originally for the soft photon emission.
In this approximation the quark energy loss distribution
has the form
\beq
P(\Delta E)=\sum_{n=0}^{\infty}\frac{1}{n!}\left[
\prod_{i=1}^{n}\int d\omega_{i}
\frac{dP(\omega_{i})}{d\omega}
\right]\delta\left(\Delta E-\sum_{i=1}^{n}\omega_{i}\right)
\exp{\left[-\int d\omega \frac{dP}{d\omega}\right]}\,,
\label{eq:70}
\eeq
where $dP/d\omega$ is the probability distribution for one gluon
emission. 
This approximation leads to the leakage 
of the probability to the unphysical region of 
$\Delta E> E$ \cite{Eskola}.
To avoid the quark charge non-conservation we define the 
renormalized distribution 
$\bar P(\Delta E)=K_{q}P(\Delta E)$ with 
$K_{q}\!=\!\int^{\infty}_{0} d\Delta E P(\Delta E)
/\!\int^{E}_{0} d\Delta E P(\Delta E)$.
We use the renormalized distribution to define the in-medium 
FF
$D_{q/q}^{ind}(z)\!=\!\bar{P}(\Delta E\!=\!E(1-z))$.
To ensure the momentum conservation 
we take into account the $q\rightarrow g$ transition as
well. At the one gluon level the corresponding FF
can be written as $D_{g/q}^{ind}(z)\!=\!dP(\omega\!=\!z E)/d\omega$. 
This automatically leads to the FFs which satisfy the
momentum sum rule. 
We use the same form of the $q\rightarrow g$
FF for the case with the multiple gluon emission.
To satisfy the momentum sum rule (which are not valid
after the renormalization of the $q\rightarrow q$ distribution)
we multiply it by a renormalization coefficient $K_{g}$
defined from the total momentum conservation. This procedure 
seems to be reasonable since
the nuclear modification factor are only sensitive the behavior of the 
FFs at $z$ close to unity \cite{BDMS_RAA} where the 
form of the $q\rightarrow g$ distribution should not be very sensitive
to the multiple gluon emission. 
In the case of the $g\rightarrow g$ transition we use the following
prescription. In the first step  
we define $D_{g/g}^{ind}$ at 
$z>0.5$ through the Landau distribution $P(\Delta E)$, and in the soft 
region $z<0.5$ (where 
the multiple gluon emission and the Sudakov suppression
strongly compensate each other) we use the one gluon distribution.
Then we multiply this FF by a renormalization coefficient
$\bar{K}_{g}$ to ensure the momentum conservation (since the number 
of gluons is not conserved the arguments based on the conservation of the
probability cannot be used in this case). 

In the above discussion we ignored the collisional energy loss.
Presently there is no an accurate method for incorporating 
of the collisional energy loss in the scheme of the medium-modified 
FFs. 
In the present work we view the collisional energy loss as a perturbation 
and incorporate it into our model
by a small renormalization of the QGP density according to the 
change in the $\Delta E$ due to the collisional energy loss.   
To evaluate 
the collisional energy loss we use the Bjorken method \cite{Bjorken1}
with an accurate
treatment of kinematics of the binary collisions (the details can be found
in \cite{Z_Ecoll}).
We use the same infrared cutoffs
and parametrization of the coupling constant for the radiative and collisional
energy loss, which is important for minimizing the theoretical 
uncertainties in the fraction  of the collisional contribution.

We calculate the cross sections for the  
$N+N\rightarrow q(g)+X$ processes using the LO 
pQCD formula with the CTEQ6 \cite{CTEQ6} parton distribution functions.
To account for the nuclear modification of the parton densities
(which leads to some small deviation of $R_{AA}$ from unity even without
parton energy loss) we include the EKS98 correction \cite{EKS98}.
To simulate the higher order $K$-factor
we follow the prescription used in the PYTHIA event generator \cite{PYTHIA}
with replacement of the $Q$ argument of $\alpha_{s}$ by a 
lower value $cQ$. We take $c=0.265$ which allows to describe 
well the data on $\pi^{0}$ $p_{T}$-spectrum in the $pp$-collisions.
For the FFs $D_{h/q(g)}(z,Q_{0})$ we use the KKP parametrization \cite{KKP}.

\vspace{.2cm}
\noindent{\bf 3.}
For calculations  of the induced gluon spectrum and collisional energy loss 
with the help of the formulas given in \cite{Z04_RAA,Z_Ecoll} we must 
specify the form of the coupling constant
and the mass parameters (the quasiparticle masses and the Debye mass).
In our calculations we use the running coupling constant.
We parametrize $\alpha_{s}(Q^{2})$ by the one-loop expression
and assume that it is frozen at some value $\alpha_{s}^{fr}$ for $Q\le Q_{fr}$. 
This form with $\alpha_{s}^{fr}\approx 0.7$ 
($Q_{fr}\approx 0.82$ GeV for $\Lambda_{QCD}=0.3$ GeV)
allows one to describe well
the HERA data on the low-$x$ structure 
functions within the dipole approach \cite{NZ,NZ_HERA,NZZ}.
A similar value of 
$\alpha_{s}^{fr}$ follows from the relation 
$
\int_{\mbox{\small 0}}^{\mbox{\small 2 GeV}}\!dQ\frac{\alpha_{s}(Q^{2})}{\pi}
\approx 0.36 \,\,
\mbox{GeV}\,
$
obtained in \cite{DKT} from the analysis of the heavy quark energy 
loss in vacuum.
In vacuum the stopping of the growth of $\alpha_{s}$  
at low $Q$ may be caused by the nonperturbative effects \cite{DKT}. 
In the QGP thermal partons can give additional suppression of 
$\alpha_{s}$ at low momenta ($Q\sim 2-3T$).
The lattice simulations \cite{alpha-lattice} give  
$\alpha_{s}(T)$ smoothly decreasing  from $\sim 0.5$ at 
$T\approx 175$ MeV to $\sim 0.35$ at $T\approx 400$ MeV.
However, the thermal $\alpha_{s}(T)$ in some sense 
gives the mean value of $\alpha_{s}$. For this reason one   
can expect that the thermal $\alpha_{s}(T)$ should be somewhat 
smaller than the in-medium $\alpha_{s}^{fr}$.
To clear up whether the RHIC data on jet quenching agree with the thermal
suppression of $\alpha_{s}$ we perform numerical calculations for different 
values of $\alpha_{s}^{fr}$.

As in \cite{Z_Ecoll} we use the quasiparticle masses
obtained in Ref. \cite{LH} from the 
analysis of the lattice data within the quasiparticle model. 
For the relevant range of the plasma temperature $T\sim (1-3)T_{c}$ 
the analysis \cite{LH} gives $m_{q}\approx 0.3$ and $m_{g}\approx 0.4$ GeV. 
To fix the Debye mass in the QGP we use the results 
of the lattice calculations for $N_{f}=2$ \cite{Bielefeld_Md} which 
give the ratio $\mu_{D}/T$ slowly decreasing with $T$  
($\mu_{D}/T\approx 3$ at $T\sim 1.5T_{c}$, $\mu_{D}/T\approx 2.4$ at 
$T\sim 4T_{c}$). 

\noindent{\bf 4.} 
We describe the QGP
in the Bjorken model \cite{Bjorken2} with the 
longitudinal expansion which gives the proper time dependence
of the plasma temperature $T^{3}\tau=T_{0}^{3}\tau_{0}$ ($T_{0}$ is the
initial plasma temperature).
To simplify the numerical calculations
for each value of the impact parameter $b$ we 
neglect the variation of $T_{0}$ in the transverse directions.
For each $b$ we
define its own effective initial 
temperature evaluated with the help of the entropy
distribution adjusted in the 
hydrodynamic analysis \cite{Heinz} of the RHIC data on the 
small-$p_{T}$ hadron spectra. 
For Au+Au collisions at $\sqrt{s}=200$ GeV
it reads \cite{Heinz}
$
{dS(\tau,\r,\bb)}/{d\r dz}=
\frac{C}{\tau(1+\alpha)}\left[
\alpha {dN_{part}}/{d\r}+(1-\alpha){dN_{coll}}/{d\r}\right]\,,
$
where $C=24$, and $\alpha=0.85$, $N_{part}$ and $N_{coll}$ 
are the number of participants and binary 
collisions evaluated in the Glauber model.  
In evaluating the entropy 
distribution we use the Woods-Saxon nucleus density 
$\rho_{A}(r)\!=\!C_{norm}/\{1+\exp[(r-c)/d]\}$ with $c=1.07A^{1/3}$ fm, and
$d=0.545$ fm.
For central Au+Au collisions
at $\sqrt{s}=200$ GeV this gives $T_{0}\approx 320$ MeV for $\tau_{0}=0.5$ fm.
It was assumed that the QGP occupies the region $r\!<\!(c\!+\!kd)$,
with $k=2$ (for $k=1$ $R_{AA}$ changes slightly).
To perform the extrapolation to the LHC energy $\sqrt{s}=5500$ GeV
we use the energy dependence of the entropy similar to the energy
dependence of the total particle rapidity density $\propto 
N_{part}\ln{(\sqrt{s}/1.5)}$ 
observed at RHIC \cite{dNdy}. It gives for central 
Pb+Pb collisions $T_{0}\approx 404$ MeV.
The fast parton path length in the QGP, $L$, 
in the medium has been calculated according to the position
of the hard reaction in the impact parameter plane.
To take into account the fact at times about $1-2$ units of 
$R_{A}$ the transverse expansion
should lead to fast cooling of the hot QCD matter \cite{Bjorken2} we also 
impose the condition $L< L_{max}$. We performed the calculations for two
values $L_{max}=6$ and 8 fm. 

\vspace{.2cm}
\noindent{\bf 5.}
We present the numerical results for $\alpha_{s}^{fr}=0.7$, 0.5 and 0.4.
The higher value seems to be reasonable in the absence
of the thermal effects since it comes from the analyses of the HERA data 
on the low-$x$ structure functions 
\cite{NZ_HERA} and heavy quark energy loss in vacuum \cite{DKT}.
In Fig.~1 we plot $R_{AA}$ for $\pi^{0}$ production in 
the central Au+Au collisions at $\sqrt{s}=200$ GeV. 
The theoretical curves corresponds to $L_{max}=8$ fm. The choice
$L_{max}=6$ fm gives $R_{AA}$ higher just by about 3-8\%.
The experimental points in Fig.~1 are from \cite{PHENIX08}. 
The upper and lower panels show the results for the chemically equilibrium 
and purely gluonic plasmas, respectively.
The results are presented for the purely radiative energy loss
and with inclusion of the collisional energy loss and the radiative energy
gain. We have found that the effect of the radiative energy gain on 
 $R_{AA}$ is practically negligible and can be safely neglected. 
Besides the total (quarks plus gluons) contribution in Fig.~1 we also show
separately the contributions from quarks and gluons.
The grows of $R_{AA}$ for gluons is due to the $q\rightarrow g$ transition
which is usually neglected. However, it does not affect strongly the total
$R_{AA}$ since for $\sqrt{s}=200$ GeV 
the gluon contribution to the hard cross section is small at $p_{T}\gsim
15$ GeV.
As can be seen from Fig.~1 the collisional energy loss suppresses  
$R_{AA}$ only by about 15-25\%.
This is in contradiction with strong collisional JQ found in \cite{Thoma}. 
To illustrate the effect of the time ordering of the DGLAP and induced
radiation stages in Fig.~1 we also plot the results for the inverse
order of these stages. One can see that the results for these two prescription
are very close. This fact may be explained by the dominance of the soft gluon
emission at RHIC energies.

From Fig.~1 one can see that the theoretical $R_{AA}$ for the chemically 
equilibrium plasma obtained with $\alpha_{s}^{fr}=0.5$ is in 
qualitative agreement with the experimental one. 
The scenario with purely gluonic plasma can be consistent
with the RHIC data if $\alpha_{s}^{fr}\approx0.4$.
In the light of the lattice results \cite{alpha-lattice} 
the value $\alpha_{s}^{fr}\sim 0.5$
seems to be reasonable for the  RHIC (and LHC) conditions.
Thus one sees that, contrary to the conclusion of Ref. \cite{BMueller},
the scenario with the chemically equilibrated plasma 
can not be excluded. Note that the distribution of the entropy \cite{Heinz}
used in our calculations gives the entropy rapidity density $dS/dy\approx
6500$. If we take a smaller value $dS/dy\approx 5100$ obtained in 
\cite{BM-entropy} approximately the same $R_{AA}$ as for 
$\alpha_{s}^{fr}=0.4$ and 0.5 can be obtained with the values 
$\alpha_{s}^{fr}\approx0.45$ and
0.55. 

In Fig.~2 we show  $R_{AA}$ as a function of $N_{part}$.
One sees that the model reproduces qualitatively the growth
of $R_{AA}$ with decrease of $N_{part}$. But for
very peripheral collisions with small $N_{part}$ it overestimates 
the observed $R_{AA}$. This may be connected with inadequacy of the neglect of
the transverse motion of the matter for thin plasma. Also, in this region
the neglect of the variation of $T_{0}$ in the impact parameter space
may be inadequate as well. Probably for similar reasons the model
underestimate the ellipticity parameter $v_{2}$ (which is also
sensitive to the evolution of the QGP for the peripheral collisions). 
Our calculations give $v_{2}\sim 0.05-0.08$ for $p_{T}\sim 5$ GeV while the
experiment gives $v_{2}\sim 0.1-0.15$ \cite{V2}.   

In Fig.~3 we plot the theoretical results similar to that shown in Fig.~1
but for Pb+Pb collisions at LHC for $\sqrt{s}=5500$ GeV. 
One can see that the effect of the collisional energy loss becomes
smaller for LHC conditions. But the effect of the time ordering of the 
DGLAP and induced radiation stages is bigger as compared to the RHIC.
The difference between the results for the chemically equilibrium and 
non-equilibrium plasmas is relatively small.

\vspace{.2cm}
\noindent {\bf 6}. In summary, we have calculated the nuclear modification
factor for the RHIC and LHC conditions accounting for both the radiative and 
collisional energy losses with the 
running $\alpha_{s}$.
The radiative energy loss has been calculated 
within the LCPI approach \cite{LCPI}. The
collisional energy loss has been evaluated in the Bjorken model
of elastic binary collisions with an accurate
treatment of kinematics of the binary collisions.
In contrast to \cite{Thoma} we find relatively small
effect of the collisional energy loss on JQ.
We also investigated the effect of the parton energy
gain due to the induced gluon absorption. We find that this effect
is negligible for the RHIC and LHC conditions.

The calculations are performed using a new algorithm
for the multiple gluon emission which takes into account the time ordering 
of the DGLAP and induced gluon emission stages. We find
that  the effect of the ordering of these two stages is 
relatively small for RHIC conditions, but becomes bigger for LHC.

Comparison of our theoretical results with the RHIC data  
show that $R_{AA}$ can be described in the scenario with the 
chemically equilibrium QGP with the entropy extracted from the 
hydrodynamical simulation of the $AA$ collisions at RHIC energies.
The scenario with the purely gluonic plasma is 
also possible, but requires somewhat stronger thermal 
suppression of $\alpha_{s}$.
This contradicts to the conclusion
of Ref. \cite{BMueller} that the observed JQ and total entropy 
of the QGP are incompatible with the chemically equilibrium plasma scenario.

\vspace {.7 cm}
\noindent
{\large\bf Acknowledgements}

\noindent
This research is supported 
in part by the grant RFBR
06-02-16078-a and the program SS-3472.2008.2.

\newpage
\vspace{-.5cm}
\begin{center}
{\Large \bf Figures}
\end{center}
\begin{figure}[ht]
\vspace{-.5cm}
\begin{center}
\epsfig{file=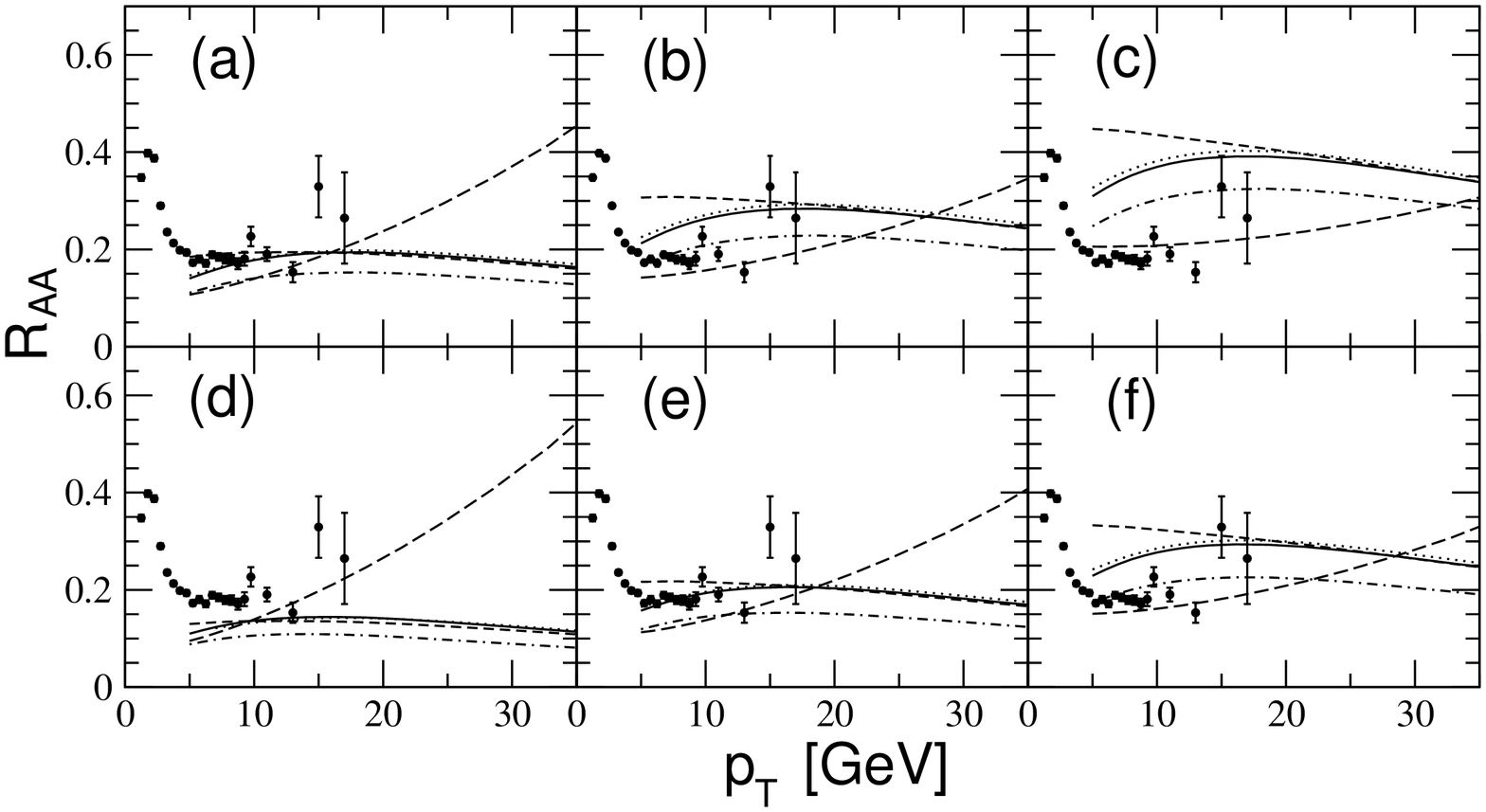,height=12cm}
\vspace{-1.5cm}
\end{center}
\caption[.]
{
The nuclear modification factor $R_{AA}$ for $\pi^{0}$ production
in the central Au+Au collisions
at $\sqrt{s}=200$ GeV for $\alpha_{s}^{fr}=0.7$ ((a),(d)),
$\alpha_{s}^{fr}=0.5$ ((b),(e)),
$\alpha_{s}^{fr}=0.4$ ((c),(f)).
The upper panels are for the chemically equilibrium
plasma, and the lower ones for purely gluonic plasma with the same entropy.
For all the theoretical curves $L_{max}=8$ fm. 
Solid line: the total (quarks plus gluons) radiative $R_{AA}$.
Dashed line: the radiative $R_{AA}$ for $\pi^{0}$ 
from quarks.
Long-dash line: the radiative $R_{AA}$ for $\pi^{0}$ from gluons.
Dash-dotted line: the total (quarks plus gluons) $R_{AA}$
including the radiative plus collisional energy loss and 
the energy gain due to gluon absorption.
Dotted curves show the total (quarks plus gluons) radiative 
$R_{AA}$ for the inverse time order 
of the DGLAP and induced gluon emission stages.
The experimental points are the data obtained by the PHENIX
Collaboration \cite{PHENIX08} for the most central (0-5\%) collisions.  
}
\end{figure}
\begin{figure}[ht]
\begin{center}
\epsfig{file=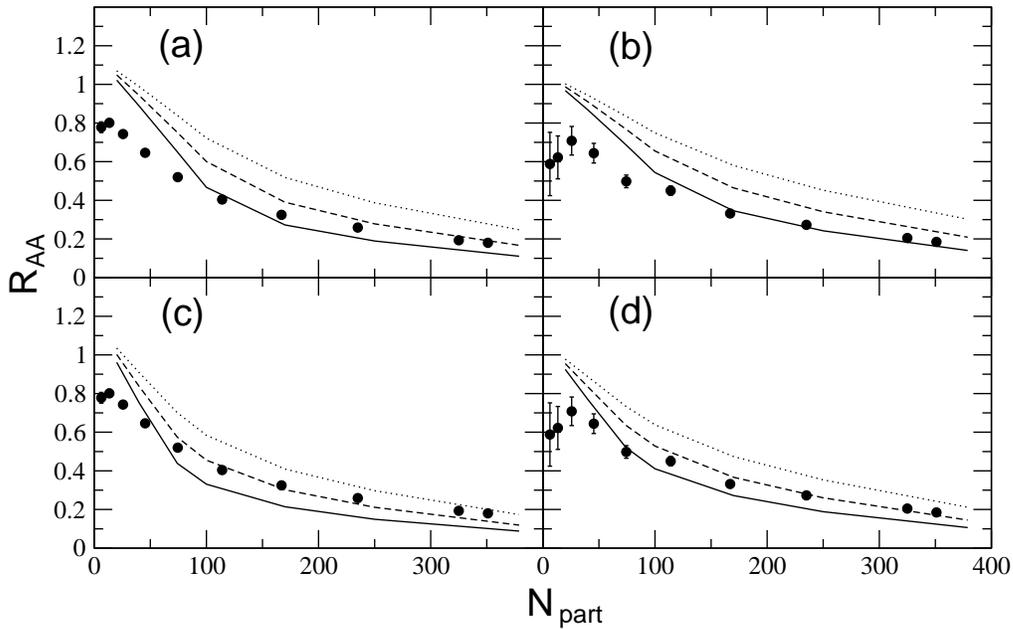,height=12cm}
\vspace{-1cm}
\end{center}
\caption[.]
{
The nuclear modification factor $R_{AA}$ for $\pi^{0}$
production in Au+Au at $\sqrt{s}=200$ GeV for
$p_{T}>5$ GeV (left panels) and $p_{T}>10$ GeV (right panels)
as a function of $N_{part}$.
The upper panels are for the chemically equilibrium
plasma, and the lower ones for purely gluonic plasma with the same entropy.
The theoretical curves show the total (quarks plus gluons) $R_{AA}$
including the radiative plus collisional energy loss and the energy gain
due to gluon absorption for $\alpha_{s}^{fr}=0.7$ (solid line),
$\alpha_{s}^{fr}=0.5$ (dashed line),
$\alpha_{s}^{fr}=0.7$ (dotted line). The experimental points are from 
\cite{PHENIX08}.
}
\end{figure}
\newpage
\begin{figure}[ht]
\begin{center}
\epsfig{file=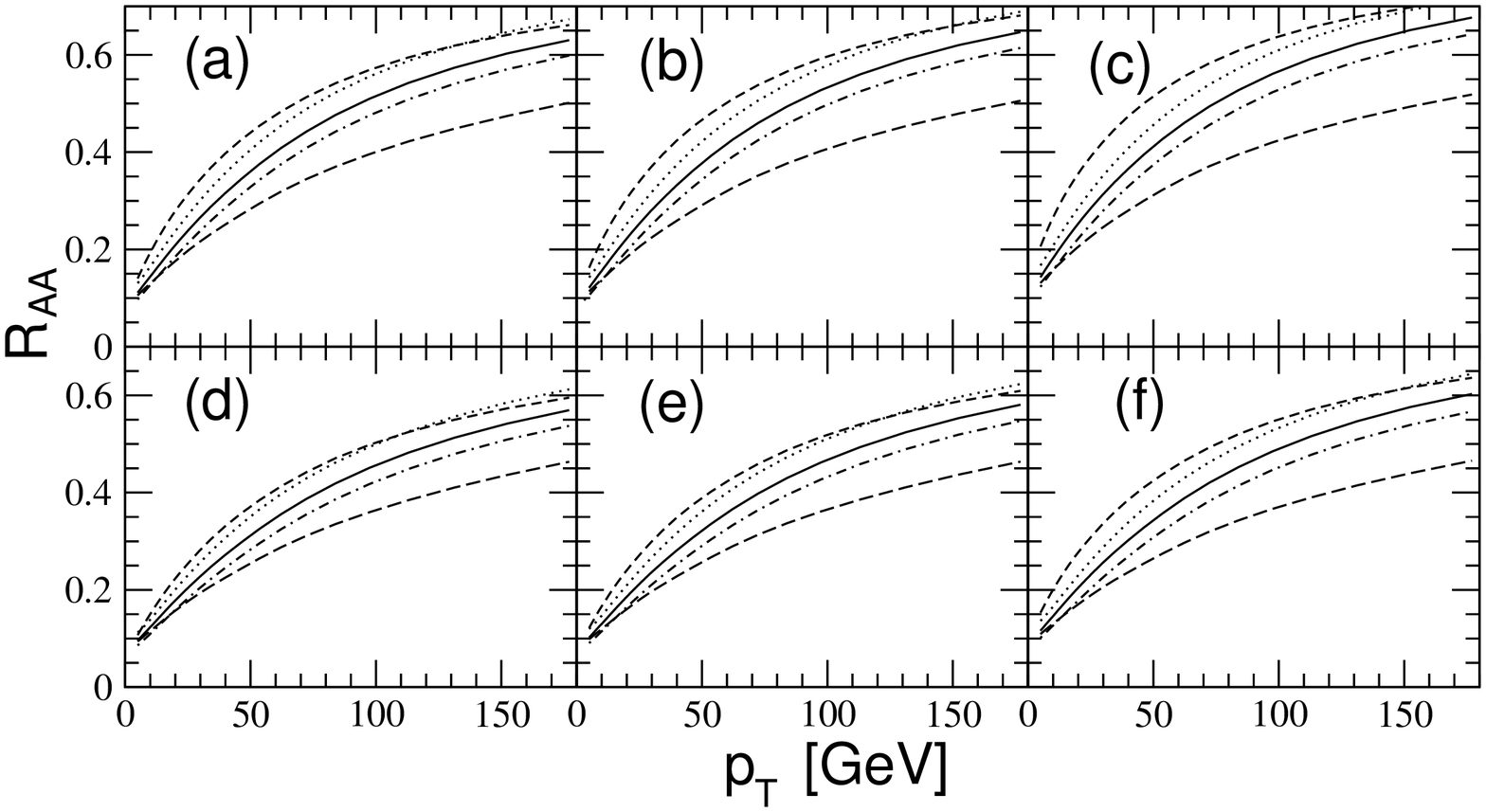,height=12cm}
\vspace{-1cm}
\end{center}
\caption[.]{
The same as in Fig.~1 for Pb+Pb collisions at $\sqrt{s}=5500$ GeV.
}
\end{figure}


\begin{thebibliography}{99}
\bibitem{RHIC_data}
P.M.~Jacobs, M. van Leeuwen,
Nucl. Phys. A{\bf 774}, 237 (2006) 
and references therein.

\bibitem{BDMPS}
R.~Baier, Y.L.~Dokshitzer, A.H.~Mueller, S.~Peign\'e, and D.~Schiff,
Nucl.\ Phys.\ B{\bf 483}, 291 (1997); {\it ibid.} B{\bf 484}, 265 (1997);
%
R.~Baier, Y.L.~Dokshitzer, A.H.~Mueller, and D.~Schiff,
Nucl.\ Phys.\ B{\bf 531}, 403 (1998).

\bibitem{LCPI}
B.G. Zakharov, JETP\ Lett. {\bf 63}, 952 (1996); {\em ibid}
{\bf 65}, 615 (1997);
{\bf 70}, 176 (1999);
Phys.\ Atom.\ Nucl. {\bf 61}, 838 (1998).


\bibitem{BSZ}
R. Baier, D. Schiff, and B.G. Zakharov, 
Ann.\ Rev.\ Nucl.\ Part. {\bf 50}, 37 (2000) [arXiv:hep-ph/0002198].

\bibitem{W1}
U.A.~Wiedemann,
Nucl.\ Phys.\ A{\bf 690}, 731 (2001).


\bibitem{GLV1}
M.~Gyulassy, P.~L\'evai and I.~Vitev, 
Nucl.\ Phys. B{\bf 594}, 371 (2001).

\bibitem{AMY}
P.~Arnold, G.D.~Moore, and L.G.~Yaffe,
JHEP {\bf 0206}, 030 (2002).

\bibitem{BDMS_RAA}
R.~Baier, Yu.L.~Dokshitzer, A.H.~Mueller, and
D.~Schiff, JHEP {\bf 0109}, 033 (2001). 



\bibitem{Eskola}
K.J.~Eskola, H.~Honkanen, C.A.~Salgado, and U.A.~Wiedemann,
Nucl. Phys. A{\bf 747}, 511 (2005).



\bibitem{Gale}
G.Y.~Qin {\it et al.}, 
Phys. Rev. C{\bf 76}, 064907 (2007).

\bibitem{Armesto}
N.~Armesto, L.~Cunqueiro, C.A.~Salgado,
and W.C. Xiang, JHEP {\bf 0802}, 048 (2008).

\bibitem{Zapp}
K.~Zapp, G.~Ingelman, J.~Rathsman, J.~Stachel, and U.A.~Wiedemann,
arXiv:0804.3568 [hep-ph].

\bibitem{Lokhtin}
I.P.~Lokhtin, {\it et al.},
arXiv:0810.2082 [hep-ph].

\bibitem{Z_OA}
B.G.~Zakharov, JETP Lett. {\bf 73}, 49 (2001).

\bibitem{AZZ}
P.~Aurenche, B.G.~Zakharov, and H.~Zaraket,
JETP Lett. {\bf 87}, 605 (2008) [arXiv:0804.4282 [hep-ph]].

\bibitem{PHENIX08}
A. Adare {\it et al.} [PHENIX Collaboration],
arXiv:0801.4020 [nucl-ex].


\bibitem{BMueller}
B.~Muller and J.L.~Nagle,
Ann. Rev. Nucl. Part. Sci. {\bf 56}, 93 (2006) [arXiv:nucl-th/0602029].


\bibitem{NZ_HERA}
N.N.~Nikolaev and B.G.~Zakharov,
Phys. Lett. B{\bf 327}, 149 (1994). 


\bibitem{PYTHIA}
T.~Sjostrand, L.~Lonnblad, S.~Mrenna, and  P.~Skands,
arXiv:hep-ph/0308153.


\bibitem{Z04_RAA}
B.G.~Zakharov, JETP Lett. {\bf 80}, 617 (2004).




\bibitem{Bjorken1} J.D.~Bjorken, Fermilab preprint 
82/59-THY (1982, unpublished).

\bibitem{Z_Ecoll}
B.G.~Zakharov,
JETP Lett. {\bf 86}, 444 (2007)
[arXiv:0708.0816 [hep-ph]].


\bibitem{CTEQ6}
S.~Kretzer, H.L.~Lai, F.~Olness, and W.K.~Tung,
Phys. Rev. D{\bf 69}, 114005 (2004).

\bibitem{EKS98}
K.J.~Eskola, V.J.~Kolhinen, and C.A.~Salgado,
Eur. Phys. J. C{\bf 9}, 61 (1999).


\bibitem{KKP}
B.~A. Kniehl, G. Kramer, and B. Potter, 
Nucl.\ Phys. B{\bf 582}, 514 (2000).

\bibitem{NZ}
N.N.~Nikolaev and B.G.~Zakharov,
Z. Phys. C{\bf 49}, 607 (1991). 

\bibitem{NZZ}
N.N.~Nikolaev, B.G.~Zakharov, and V.R.~Zoller,
Phys. Lett. B{\bf 328}, 486 (1994).


\bibitem{DKT}
Yu.L.~Dokshitzer, V.A.~Khoze, and S.I.~Troyan,
Phys.\ Rev. D{\bf 53}, 89 (1996).
 

\bibitem{alpha-lattice} O.~Kaczmarek, F.~Karsch, F.~Zantow, and 
P.~Petreczky,
Phys. Rev. D{\bf 70}, 074505 (2004).

\bibitem{LH}
P.~L\'evai and U.~Heinz,
Phys.\ Rev.\ C{\bf 57}, 1879 (1998).

\bibitem{Bielefeld_Md}
O.~Kaczmarek and F.~Zantow,
Phys. Rev. D{\bf 71}, 114510 (2005).

\bibitem{Bjorken2}
J.D.~Bjorken, 
Phys.\ Rev. D{\bf 27}, 140 (1983).


\bibitem{Heinz}
T.~Hirano, U.W.~Heinz, D.~Kharzeev, R.~Lacey, and Y.~Nara,
Phys. Lett. B{\bf 636}, 299 (2006).

\bibitem{dNdy}
S.S.~Adler {\it et al.} [PHENIX Collaboration],
Phys. Rev. C{\bf 71}, 034908 (2005).

\bibitem{Thoma}
M.G.~Mustafa and M.H.~Thoma,
Acta Phys. Hung. A{\bf 22}, 93 (2005)
[arXiv:hep-ph/0311168].


\bibitem{BM-entropy}
B.~Muller and K.~Rajagopal,
Eur. Phys. J. C{\bf 43}, 15 (2005).


\bibitem{V2}
S.S.~Adler {\it et al.} [PHENIX Collaboration],
Phys. Rev. Lett. {\bf 96}, 032302 (2006).

\end{thebibliography}
\end{document}